\documentclass[prl,amsmath,amssymb,twocolumn,superscriptaddress]{revtex4}

\usepackage{graphicx}
\usepackage{latexsym}
\usepackage{dcolumn}

\usepackage{bm}

\usepackage{bm}

\begin{document}

\title{Direct mapping of the spin-filtered surface bands of a three-dimensional quantum spin Hall insulator }

\author{Akinori Nishide}
\address{Institute for Solid State Physics, the University of Tokyo, Japan}
\author{Alexey A. Taskin}
\address{Institute of Scientific and Industrial Research, Osaka University, Ibaraki, Osaka 567-0047, Japan}
\author{Yasuo Takeichi}
\address{Institute for Solid State Physics, the University of Tokyo, Japan}
\author{Taichi Okuda}
\address{Institute for Solid State Physics, the University of Tokyo, Japan}
\author{Akito Kakizaki}
\address{Institute for Solid State Physics, the University of Tokyo, Japan}
\author{Toru Hirahara}
\address{Department of Physics, School of Science, the University of Tokyo, Japan}
\author{Kan Nakatsuji}
\address{Institute for Solid State Physics, the University of Tokyo, Japan}
\author{Fumio Komori}
\address{Institute for Solid State Physics, the University of Tokyo, Japan}
\author{Yoichi Ando}
\email{y_ando@sanken.osaka-u.ac.jp}
\address{Institute of Scientific and Industrial Research, Osaka University, Ibaraki, Osaka 567-0047, Japan}
\author{Iwao Matsuda}
\email{imatsuda@issp.u-tokyo.ac.jp}
\address{Institute for Solid State Physics, the University of Tokyo, Japan}

\date{\today}


\begin{abstract}
Spin-polarized band structure of the three-dimensional quantum spin Hall insulator $\rm Bi_{1-x}Sb_{x}$ (x=0.12-0.13) was fully elucidated by spin-polarized angle-resolved photoemission spectroscopy using a high-yield spin polarimeter equipped with a high-resolution electron spectrometer. Between the two time-reversal-invariant points, $\bar{\varGamma}$ and $\bar{M}$, of the (111) surface Brillouin zone, a spin-up band ($\Sigma_3$ band) was found to cross the Fermi energy only once, providing unambiguous evidence for the strong topological insulator phase. The observed spin-polarized band dispersions determine the ``mirror chirality" to be -1, which agrees with the theoretical prediction based on first-principles calculations. 
\end{abstract}

\maketitle

The spin Hall effect, which makes it possible to produce spin currents without magnet, has recently attracted a lot of attention for its potential impact on future spintronics.\cite{MurakamiScience,SinovaPRL} The spin Hall effect has also stimulated physicists not only to understand the intriguing phenomena, but also to extend the theoretical frameworks to the ``quantum spin Hall" (QSH) effect \cite{Zhang06,HaldanePRL06,MurakamiPRL06PRB08,MurakamiPRL06PRB08p2}, which is realized in a topologically non-trivial electronic state, as in the case of the quantum Hall effect. The theories of QSH effect have been constructed for two- and three-dimensions, and experimentalists have already attempted to obtain evidence for those topologically non-trivial states of matter. The QSH phase in two-dimensions (2D) is gapped in the bulk but is gapless for edge modes . Unlike the spin itself, the spin current is time-reversal invariant. Hence these modes carry spin currents without breaking time-reversal symmetry. Furthermore, the edge modes are robust against disorder or modest changes of boundary conditions, such as surface roughness or impurities. The $Z_2$ topological number $\nu $, which represents the number of Kramers pairs in the edge states, characterizes this topological protection \cite{KanePRL05,KanePRL05p2} in 2D and distinguishes the topological insulator ($\nu = 1$) from an ordinary insulator ($\nu = 0$). This 2D QSH phase has been theoretically proposed to be realized in bismuth bilayers \cite{MurakamiPRL06PRB08} and in CdTe/HgTe/CdTe quantum wells.\cite{ZhangSc06} In the latter case, the edge states were indeed observed in recent transport experiments.\cite{KonigSc07}

In three dimensions (3D), there are four $Z_2$ invariants ($\nu_0;\nu_1 \nu_2 \nu_3$), representing time-reversal-invariant band structures.\cite{KanePRL05,KanePRL05p2} When ($\nu_0;\nu_1 \nu_2 \nu_3$) = (1;111), the system is a 3D strong topological insulator, where the ``edge states" (i.e.  2D surface states) form a Kramers pair of opposite spin currents flowing on a surface [Fig. 1(a)] and are robust against disorder.\cite{KanePRB07PRL07,KanePRB07PRL07p2,MurakamiNJP07} The spin lies within the surface plane and is perpendicular to the momentum (wavenumber) of the electron. The existence of such spin-current pairs is in sharp contrast to the trivial insulator where ($\nu_0;\nu_1 \nu_2 \nu_3$) = (0;000).  Recently, the semiconductor alloy $\rm Bi_{1-x}Sb_{x}$ (x $\sim$ 0.1) was predicted\cite{KanePRB07PRL07} to be a strong topological insulator, and subsequently a spin-integrated photoemission study has mapped the (111) surface states of this system at x=0.10 to find a trace of the predicted topological band structure.\cite{HasanNature08} However, since the surface states of topological insulators are expected to be spin-polarized, the spin characters of all the edge states in the surface Brillouin zone would need to be clarified for a complete proof of the 3D QSH phase. On the theoretical front, predictions for the surface states of Bi$_{1-x}$Sb$_x$ based on a tight-binding model and first-principles calculations differ in a fundamental way, giving different ``mirror chirality" $\eta$ \cite{KanePRB08}; since the existing data for the surface state of Bi$_{1-x}$Sb$_x$ \cite{HasanNature08} fail to elucidate $\eta$, its experimental determination is strongly called for.  Therefore, it is important to conduct direct measurements of spin-polarized surface-state bands on the $\rm Bi_{1-x}Sb_{x}$ crystals by spin- and angle-resolved photoemission spectroscopy (SARPES). In the present work, we took advantage of our new spin-polarized photoemission spectrometer with a high spin-polarimetry efficiency and a high energy resolution \cite{OkudaRSI}, and determined all the spin-polarized bands of $\rm Bi_{1-x}Sb_{x}$ (x = 0.12--0.13). Our results provide direct evidence for the QSH phase in the 3D topological insulator and clarifies the mirror chirality $\eta$, settling the essential topological structure of the surface states of this material.

\begin{figure}[htb]
\includegraphics[width=8.0cm]{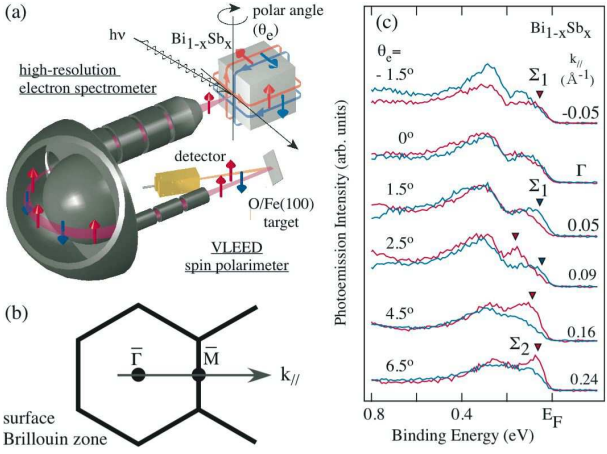}
\caption{(a) Schematic drawing of the high-resolution spin- and angle-resolved photoemission spectroscopy (SARPES) measurements. Spin-polarized (SP) electrons are depicted as up (red) and down (blue) arrows. Spin currents are also depicted for the edge states in the 3D quantum spin-Hall phase, from which the SP photoelectrons are emitted into a high-resolution electron spectrometer and the high-yield VLEED spin polarimeter. (b) The (111) surface Brillouin zone of $\rm Bi_{1-x}Sb_x$ with high symmetry points and the direction of the wavenumber $k_{//}$, which indicates the angle-scanning direction in the present photoemission experiment. (c) SARPES spectra, $\rm I_{\uparrow}$ (colored in red) and  $\rm I_{\downarrow}$ (colored in blue), of a $\rm Bi_{1-x}Sb_x$ (x=0.12) crystal at various emission angle $\theta_e$ and corresponding $k_{//}$ values. Peak positions of the surface states, $\Sigma_1$ and $\Sigma_2$, are indicated by triangles. }
\label{fig1} 
\end{figure}

$\rm Bi_{1-x}Sb_{x}$ (x=0.12-0.13) crystals were grown from a stoichiometric mixture of 99.9999$\%$ purity Bi and Sb elements by a zone melting method. The temperature dependence of the electrical resistivity indicated an opening of the bulk gap and was consistent with the published data in the literature for similar x values \cite{HasanNature08}. The crystals were cleaved along the (111) plane at 123 K below $\rm 3 \times 10^{-8}$ Pa. The SARPES were performed at BL-19A (KEK-PF, JAPAN) with a high-resolution hemispherical analyzer (SPECS Phoibos-150) equipped with a homemade high-yield spin-polarimeter using spin-dependent very-low-energy electron diffraction (VLEED).\cite{OkudaRSI} The measurement set-up is schematically drawn in Fig. 1(a). SARPES-spectra were recorded with energy and angle resolutions of 50 meV and $\rm \pm 1^\circ$, respectively. In the VLEED detector, the spin polarization P is acquired from the intensities of the reflected electrons interacted with the majority-spin (minority-spin) states of the Fe(001)p(1$\rm\times$1)-O target, $I_{maj}$ ($I_{min}$), by $\rm P=(1 / S_{eff})\times[(I_{maj}-I_{min})/(I_{maj}+I_{min})]$. Here, $\rm S_{eff}$ is the effective Sherman function of $\sim 0.2 $ determined by the reference data of a Bi sample. Use of the ferromagnetic Fe crystal surface covered with protecting oxide atomic layer enables stable spin detections with high-yields. From $I_{maj}$ and $I_{min}$, the spin-up and spin-down spectra are obtained by $\rm I_{\uparrow,\downarrow} =(I_{maj}+I_{min})(1\pm P)/2$. For comparison, spin-integrated ARPES spectra were measured with a hemispherical analyzer (VG Scienta SES-100) at energy and angle resolutions of 30 meV and $\rm \pm 0.2^\circ$. All the ARPES spectra were taken at 120-140 K with He I$\alpha$ sources along the $\bar{\varGamma}-\bar{M}$ line in the (111) surface Brillouin zone shown in Fig. 1(b). Figure 1(c) shows typical SARPES spectra of $\rm Bi_{1-x}Sb_{x}$ (x=0.12) at various wavenumbers $k_{//}$, converted from the emission angle $\theta_e$, where spin-dependent spectral features can be distinctively seen in the $\rm I_{\uparrow}$ (red) and $\rm I_{\downarrow}$ (blue) plots. Photoemission peaks assigned to the surface states are located at binding energies $E_B$ of less than 0.2 eV and are indicated by triagles. Labeling of the surface states in Fig. 1(c) were adopted from Ref. \cite{KanePRB08}.

\begin{figure}[htb]
\includegraphics[width=8.0cm]{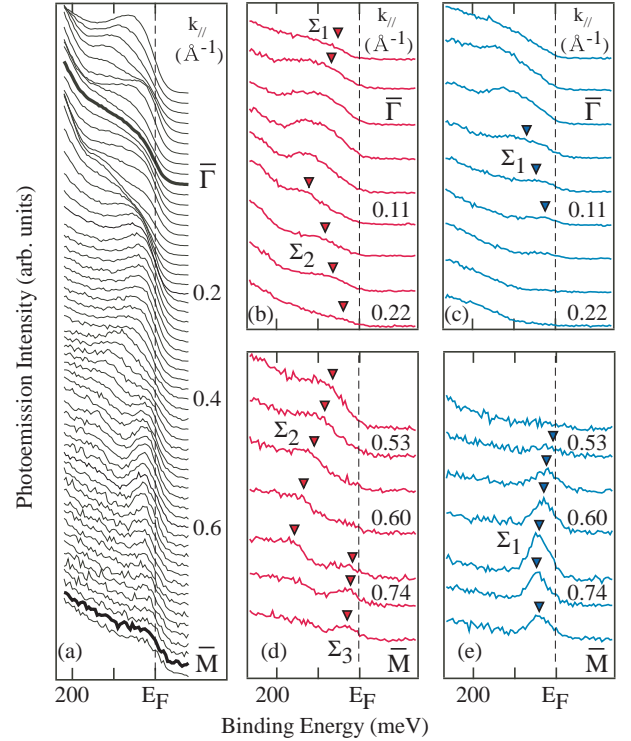}
\caption{(a) A sequence of spin-integrated ARPES spectra of $\rm Bi_{x}Sb_{1-x}$ (x=0.13) along the $\bar{\varGamma}-\bar{M}$ line. (b,c) SARPES spectra of (b) $\rm I_{\uparrow}$ (colored in red) and (c) $\rm I_{\downarrow}$ (colored in blue) for $\rm Bi_{x}Sb_{1-x}$ (x=0.12) near $\bar{\varGamma}$ . (d,e) SARPES spectra of (d) $\rm I_{\uparrow}$ and (e) $\rm I_{\downarrow}$ near $\bar{M}$. Peak positions of the surface state, $\Sigma_1$, $\Sigma_2$, and $\Sigma_3$, are indicated by triangles. 
 }
\label{fig2} 
\end{figure}

Figure 2(a) shows the spin-integrated ARPES spectra near $E_{\rm F}$ along the $\bar{\varGamma}-\bar{M}$ line. Dispersions of the surface-state bands are clearly resolved as in the previous report.\cite{HasanNature08} Figures 2(b-e) exhibit the SARPES spectra of (b,d) $\rm I_{\uparrow}$ and (c,e) $\rm I_{\downarrow}$ for each band near the $\bar{\varGamma}$ and $\bar{M}$ points. The $\Sigma_1$ band, dispersing from the $\bar{\varGamma}$ point, crosses $E_{\rm F}$ at around $k_{//}=0.1 \rm \AA^{-1}$, indicating a metallic nature. The photoemission intensity of the $\Sigma_1$ band is observed with positive wavenumbers in the $\rm I_{\downarrow}$ spectra in Fig. 2(c) and, therefore, $\Sigma_1$ has the spin-down polarization. On the other hand, the $\Sigma_2$ band, which shows up in the $\rm I_{\uparrow}$ spectra [Fig. 2(b)] and thus has the spin-up polarization, approaches $E_{\rm F}$ from $\bar{\varGamma}$ and it disperses back as $k_{//}$ increases towards $\bar{M}$ as found in Fig.2 (a). In the $\rm I_{\downarrow}$ data in Fig. 2(e), a band appears at $E_{\rm F}$ around $0.52 \rm \AA^{-1}$ and it disperses slightly to higher $E_B$. Since this band has the spin-down polarization, it is likely assigned to $\Sigma_1$. Near the $\bar{M}$ point, the $\Sigma_3$ band crosses $E_{\rm F}$ in the $\rm I_{\uparrow}$ spectra in Fig. 2(d). These two metallic bands with opposite spin polarizations, $\Sigma_1$ of down-spin and $\Sigma_3$ of up-spin, converge at the $\bar{M}$ point.

\begin{figure}[htb]
\includegraphics[width=8.0cm]{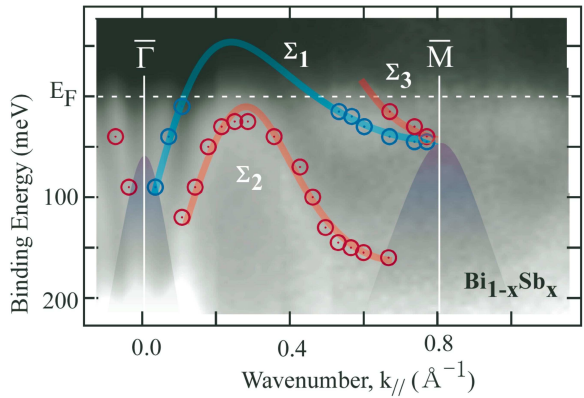}
\caption{Spin-polarazied band dispersions of the surface states of $\rm Bi_{1-x}Sb_{x}$ (x=0.12-0.13). The spin-resolved data are plotted with open circles and overlaid on a spin-integrated gray-scale diagram. Spin-up (spin-down) bands are colored in red (blue) and their dispersion curves are schematically traced by thick lines. The bulk band projection, tentatively estimated from the ARPES data, is shown as the shaded purple area.
 }
\label{fig3} 
\end{figure}

Figure 3 shows a dispersion plot of the spin-polarized surface-state bands obtained from the SARPES spectra in Fig. 2(b-e). For comparison, the data are displayed on a gray-scale band diagram constructed from the spin-integrated ARPES spectra shown in Fig. 2(a). While the band structure is essentially similar to the one reported previously \cite{HasanNature08}, it is now found to be basically composed of three surface-state bands, $\Sigma_1$, $\Sigma_2$, and $\Sigma_3$. The $\Sigma_1$ band, creating an electron pocket at the $\overline{\varGamma}$ point with the Fermi wavenumber $k_F$ of $\rm 0.1 \rm \AA^{-1}$, has asymmetric spin-polarization with respect to $\overline{\varGamma}$. The $\Sigma_2$ band disperses towards $E_{\rm F}$ and returns to higher $E_B$ without showing clear $E_{\rm F}$ crossing. Namely, in the present data, the top of the $\Sigma_2$ band likely lies below $E_{\rm F}$; this is slightly different from the previous result \cite{HasanNature08}, in which the $\Sigma_2$ band obviously crossed $E_{\rm F}$ and formed a small hole pocket. The electron filling of the surface-state bands seems to be sensitive to a slight difference of the Sb concentration (x = 0.10 in Ref. [14], while x = 0.12 -0.13 in the present work). The $\Sigma_1$ band appears again at $E_{\rm F}$ near the midpoint of the $\bar{\varGamma}-\bar{M}$ line and forms an electron pocket around $\bar{M}$. The dispersion of the $\Sigma_3$ band is very close to that of the $\Sigma_1$ band and the two bands are hardly distinguishable in the spin-integrated spectra, as shown in the grayscale diagram in Fig. 3. However, these two bands have opposite spin polarizations and, therefore, their dispersion can be unambiguously resolved in the spin-revolved spectra and plotted as the two spin-polarized bands in Fig. 3. While the previous photoemission work \cite{HasanNature08} reported a degeneracy of the $\Sigma_1$ and $\Sigma_3$ bands at $E_{\rm F}$, they actually intersect with $E_{\rm F}$ at different wavenumbers, $k_F \sim 0.52 \rm \AA^{-1}$ for $\Sigma_1$ and $k_F \sim 0.65 \rm \AA^{-1}$ for $\Sigma_3$ (Fig. 3). The $\Sigma_3$ band also creates an electron pocket around $\bar{M}$ and it converges with $\Sigma_1$ at $\bar{M}$, which is natural because these two bands have opposite spin polarities and the spin polarization P must disappear at the time-reversal-invariant point $\bar{M}$. Those spin properties demonstrated in Fig. 3 are typical for a compound with a large spin-orbit coupling, whose surface states are generally spin-split due to the loss of space inversion symmetry (Rashba effect). The Kramers' theorem dictates that the spin-split states become degenerate at the symmetry points, $\bar{\varGamma}$ and $\bar{M}$, and the spin structure naturally becomes antisymmetric with respect to these points.

Based on the $Z_2$ topology, band structures of bulk insulators are classified into trivial and non-trivial cases by examining the parity of the valence bands. Furthermore, when the system is non-trivial, it has been shown that counting the number of surface-band crossings at $E_{\rm F}$ gives a convenient way to distinguish a strong topological insulator from a weak one: odd and even number of surface-band crossings correspond to the strong and weak cases, respectively. The clear spin-polarized band structure shown in Fig. 3 signifies three $E_{\rm F}$ crossings (odd number of times) for the spin-split surface states. Consequently, one can unambiguously conclude that $\rm Bi_{1-x}Sb_{x}$ (x=0.12-0.13) is a strong topological insulator, as predicted theoretically.\cite{KanePRB07PRL07} It should be noted that the previous spin-integrated experiment \cite{HasanNature08} reached the same conclusion, though the number of the $E_{\rm F}$ crossings was different. The present study not only disentangled the complicated band structure suggested in Ref. \cite{HasanNature08}, but also unambiguously elucidated which of the many bands represents the topologically non-trivial nature; in the present case, it is the $\Sigma_3$ band that makes $\rm Bi_{1-x}Sb_{x}$ a strong topological insulator.

Now let us discuss the implications of the present results. Two theoretical approaches\cite{LiuAllenTB,GolinPR} have been employed for calculating the surface band structure of $\rm Bi_{1-x}Sb_{x}$. One was based on first-principles calculations, but was restricted to a slab or a thin-film geometry, so that the result left some ambiguity regarding the size effect.\cite{GolinPR} The other was a tight-binding model for a semi-infinite system, but the results strongly depended on the phenomenological band parameters.\cite{LiuAllenTB} These two approaches were both applied to the Bi(111) surface and yielded incompatible surface spin band structures.\cite{KanePRB08,Zhangcondmat} A recent SARPES experiment confirmed the spin-polarized surface band dispersions predicted by the first-principles calculations\cite{HiraharaPRB07}, which suggested that the former approach is likely to be appropriate for pure Bi, but the situation has not been clear for $\rm Bi_{1-x}Sb_{x}$ in the QSH phase. Recently, partly to address this issue, new topological invariants, mirror Chern number and mirror chirality, have been introduced \cite{KanePRB08}, which is analogous to the spin Chern number defined in the 2D QSH state in graphene.\cite{KanePRL05,KanePRL05p2} The main distinction in the band structures predicted by the two approaches is the presence of a crossing point (Dirac point) of the spin-split surface- bands, $\Sigma_1$ and $\Sigma_2$, that are filled at the two symmetry points, $\bar{\varGamma}$ and $\bar{M}$.\cite{KanePRB08} The mirror chirality $\eta$, which is derived from the mirror Chern number, is +1 for the crossing case (tight-binding model) and -1 for the non-crossing case (first-principles calculations).\cite{KanePRB08} As is clear in Fig. 3, the $\Sigma_1$ and $\Sigma_2$ bands do not cross each other, which leads to the conclusion that $\eta$=-1 and the first-principles approach is likely to be appropriate. Intriguingly, this implies that the $g$ factor is negative in the 3D QSH phase of $\rm Bi_{1-x}Sb_{x}$, while it is positive in pure Bi \cite{KanePRB08}. 

In conclusion, our high-resolution SARPES study elucidated the spin-polarized surface-band structures of $\rm Bi_{1-x}Sb_{x}$ (x=0.12-0.13), giving unambiguous evidence that this system is a 3D strong topological insulator. Furthermore, we were able to determine the mirror chirality of this system to be -1, which excludes the existence of a Dirac point in the middle of the $\bar{\varGamma}-\bar{M}$ line. The measured surface states are the edge-states of the 3D quantum spin Hall phase, providing robust spin-current pairs at the surface. The SARPES measurement with energy resolution below 50 meV is one of the critical techniques for complementing the topological band theory for spins and spin currents.

Shuichi Murakami is gratefully acknowledged for valuable comments and discussions. This work was partly supported by JSPS (KAKENHI 18360018, 19340078, 13674002 and 2003004) and by AFOSR (AOARD-08-4099).

\end{document}